\documentclass[aps,preprint,showpacs,groupedaddress,amsmath,amssymb]{revtex4-1}

\usepackage{graphicx}
\usepackage{dcolumn}
\usepackage{bm}

\begin{document}

\preprint{}

\title{Rotation of the polarization vector from distant radio galaxies in the perturbed FRW metric}


\author{Sankha Subhra Chakrabarty}
\email[]{s.chakrabarty@ufl.edu}
\affiliation{Department of Physics, University of Florida, Gainesville, USA}


\date{\today}

\begin{abstract}
Analysis of the correlation between the angular positions of distant radio galaxies on the sky and the orientations of their polarization vectors with respect to their major axes, indicates a dipolar anisotropy in the large scale. We consider a single mode of large-scale scalar perturbation to the FRW metric. Using Newman-Penrose formalism, we calculate the rotation of the galaxy major axis with respect to the polarization vector as the elliptic image and the polarization vector are carried through the perturbed space-time. The dependence of the rotation on the polar angular coordinate of the galaxy, is qualitatively similar to the claimed dipole pattern.
\end{abstract}


\maketitle


\section{Introduction}

\begin{figure}
\includegraphics{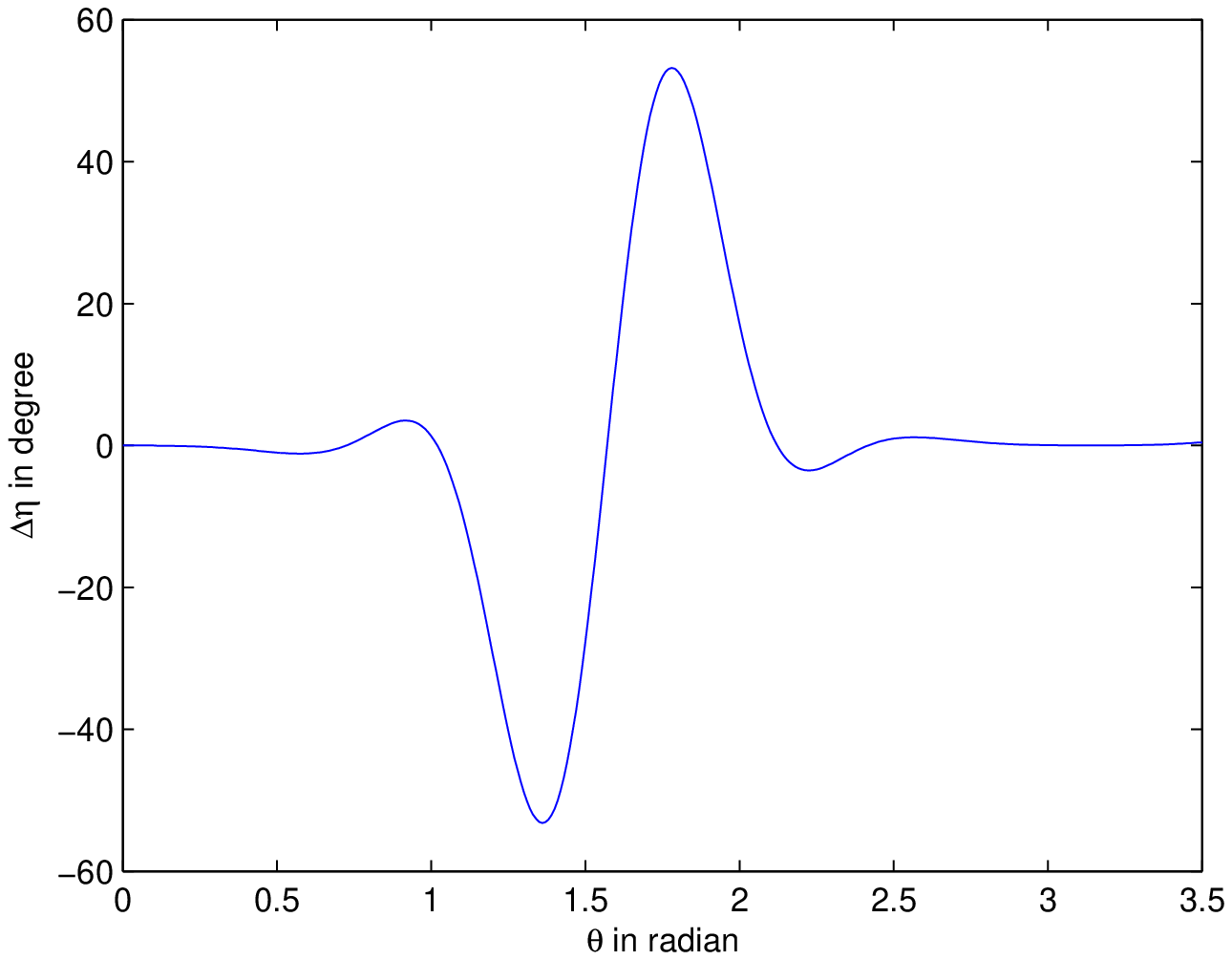}
\caption{\label{thetasine}Plot of $\Delta\eta$ (due to $\sin \vec{k}.\vec{r}$) vs $\theta$ with $\eta_{G}= 45^{\circ}, \Psi_{0} = 0.05, kr = 12$ in Eq.~(\ref{deltaetafinal}). The maximum occurs at $\theta = 1.36$ and $\Delta \eta_{max} = 53.19^{\circ}$}
\end{figure}

\begin{figure}
\includegraphics{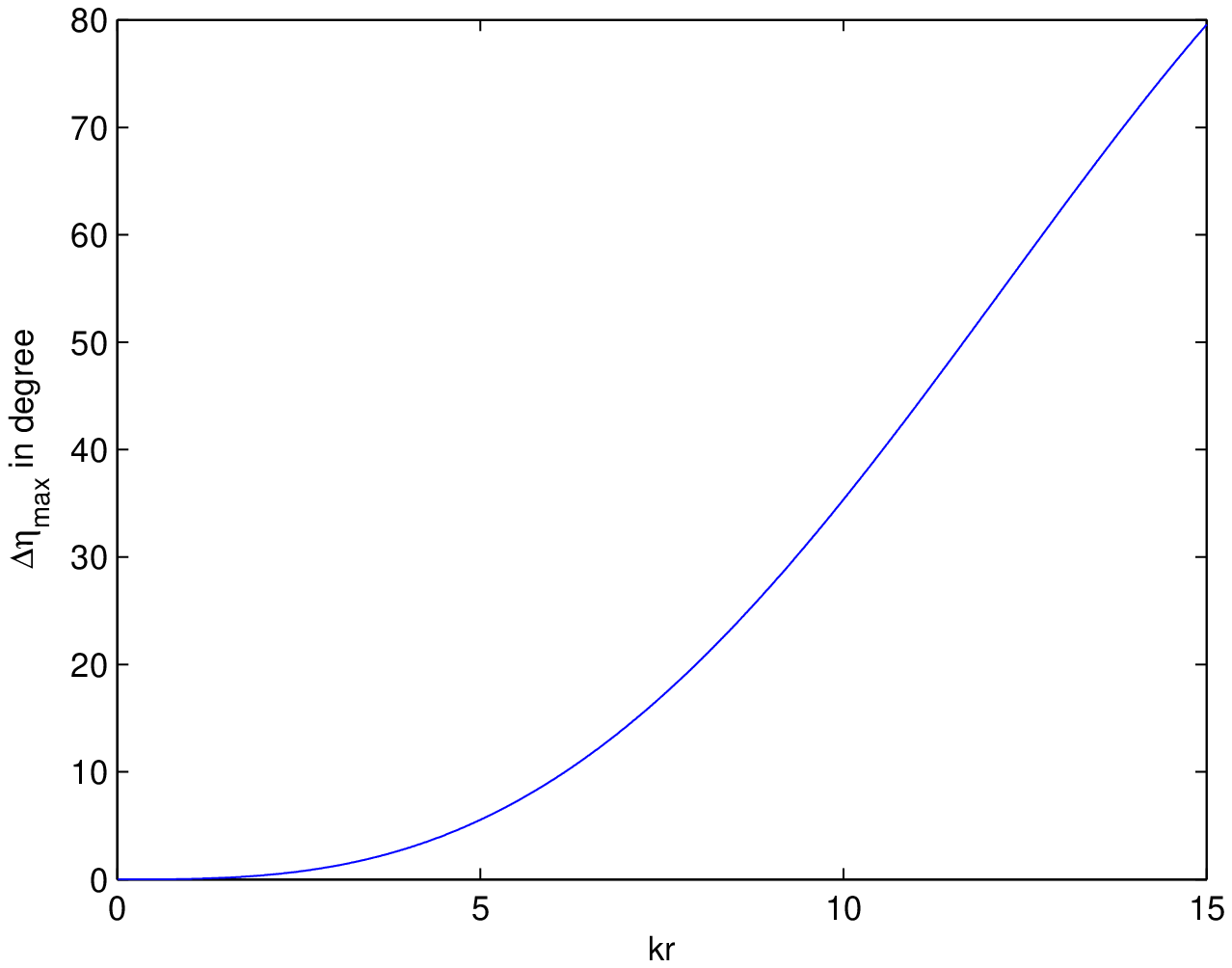}
\caption{\label{krsine}Plot of the maximum of $\Delta\eta$ (due to $\sin \vec{k}.\vec{r}$) vs $kr$ with $\eta_{G}= 45^{\circ}, \Psi_{0} = 0.05, \theta = 1.36$ in Eq.~(\ref{deltaetafinal}).}
\end{figure}

\begin{figure}
\includegraphics{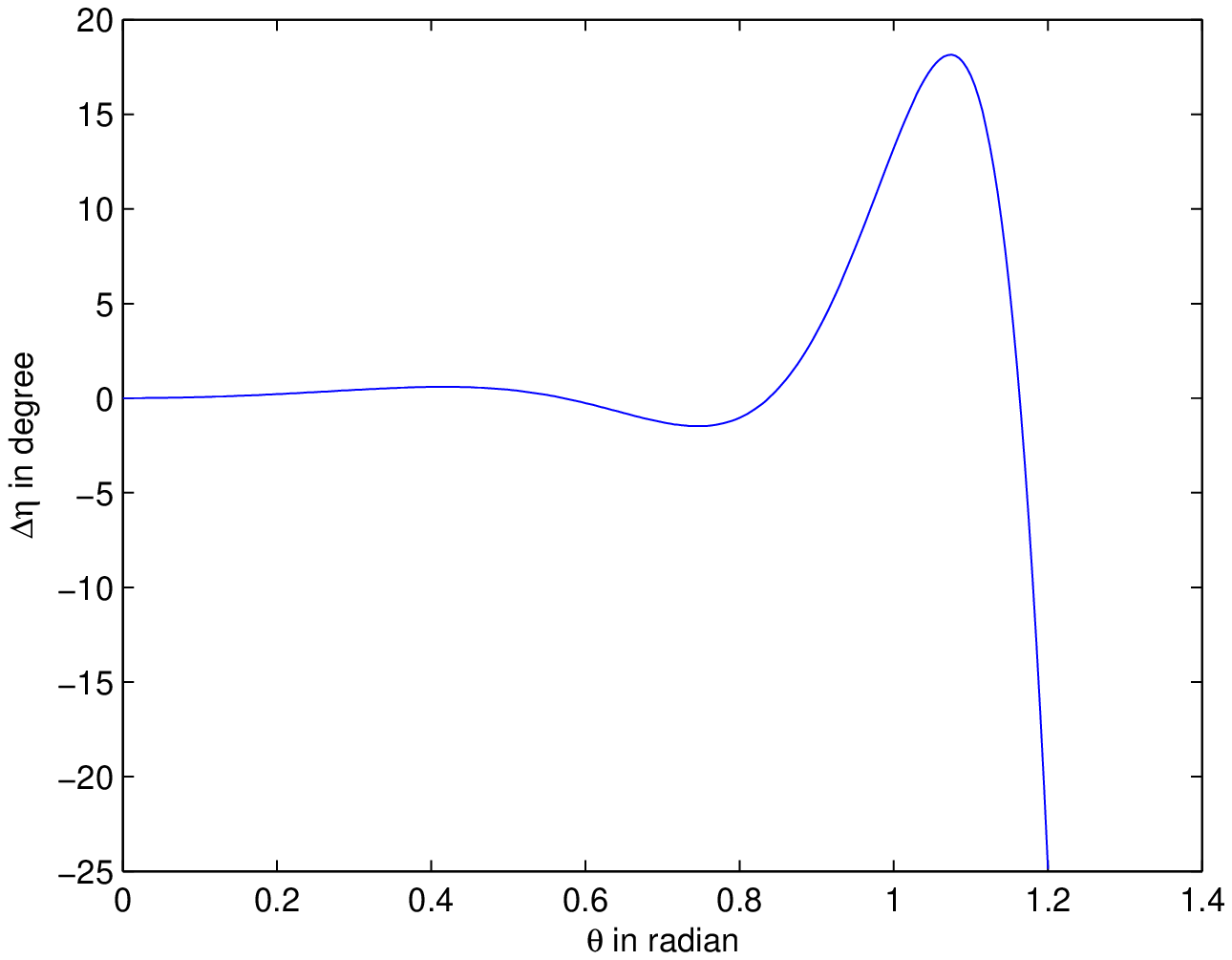}
\caption{\label{thetacos1}Plot of $\Delta\eta'$ (due to $\cos \vec{k}.\vec{r}$) vs $\theta$ for $\theta < \frac{\pi}{2}$, with $\eta_{G}= 45^{\circ}, \Psi_{0} = 0.05, kr = 12$ in Eq.~(\ref{deltaetaprime}). The maximum occurs at $\theta = 1.07$ and $\Delta \eta'_{max} = 18.13^{\circ}$}
\end{figure}

\begin{figure}
\includegraphics{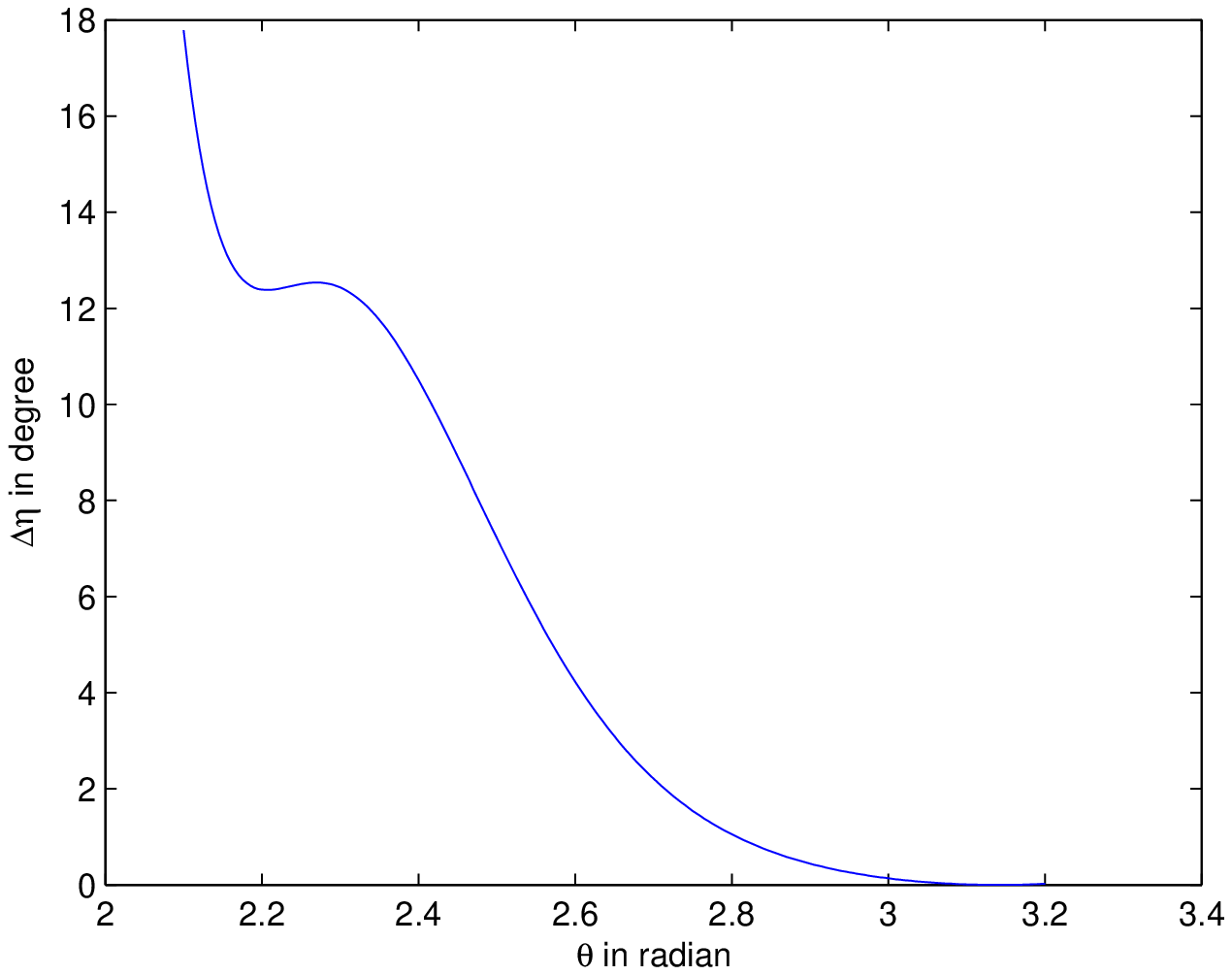}
\caption{\label{thetacos2}Plot of $\Delta\eta'$ (due to $\cos \vec{k}.\vec{r}$) vs $\theta$ for $\theta > \frac{\pi}{2}$, with $\eta_{G}= 45^{\circ}, \Psi_{0} = 0.05, kr = 12$ in Eq.~(\ref{deltaetaprime}). The maximum occurs at $\theta = 2.27$ and $\Delta \eta'_{max} = 12.54^{\circ}$}
\end{figure}

\begin{figure}
\includegraphics{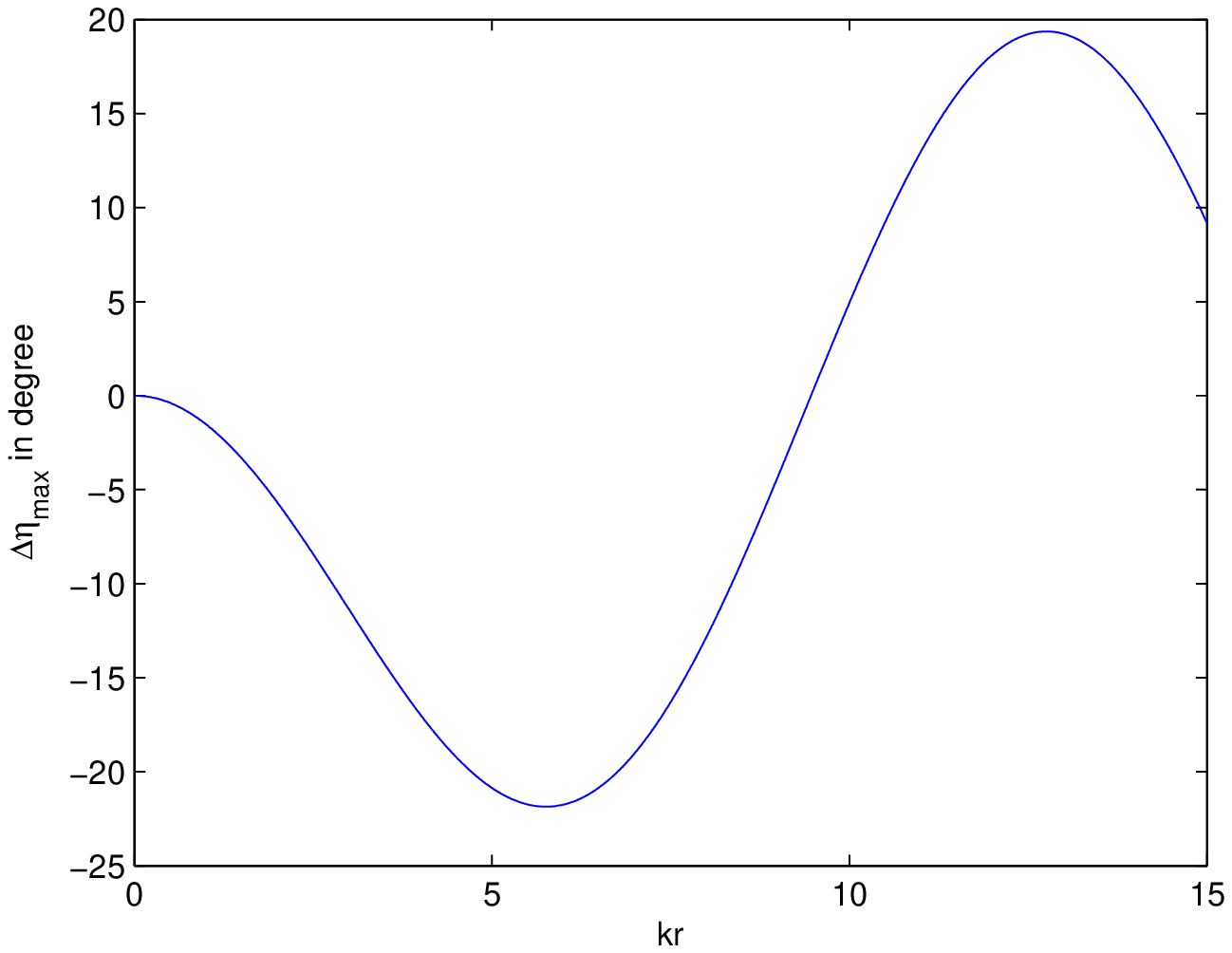}
\caption{\label{krcos1}Plot of the maximum of $\Delta\eta'$ (due to $\cos \vec{k}.\vec{r}$) vs $kr$ with $\eta_{G} = 45^{\circ}, \Psi_{0} = 0.05, \theta = 1.07$ in Eq.~(\ref{deltaetaprime}).}
\end{figure}

\begin{figure}
\includegraphics{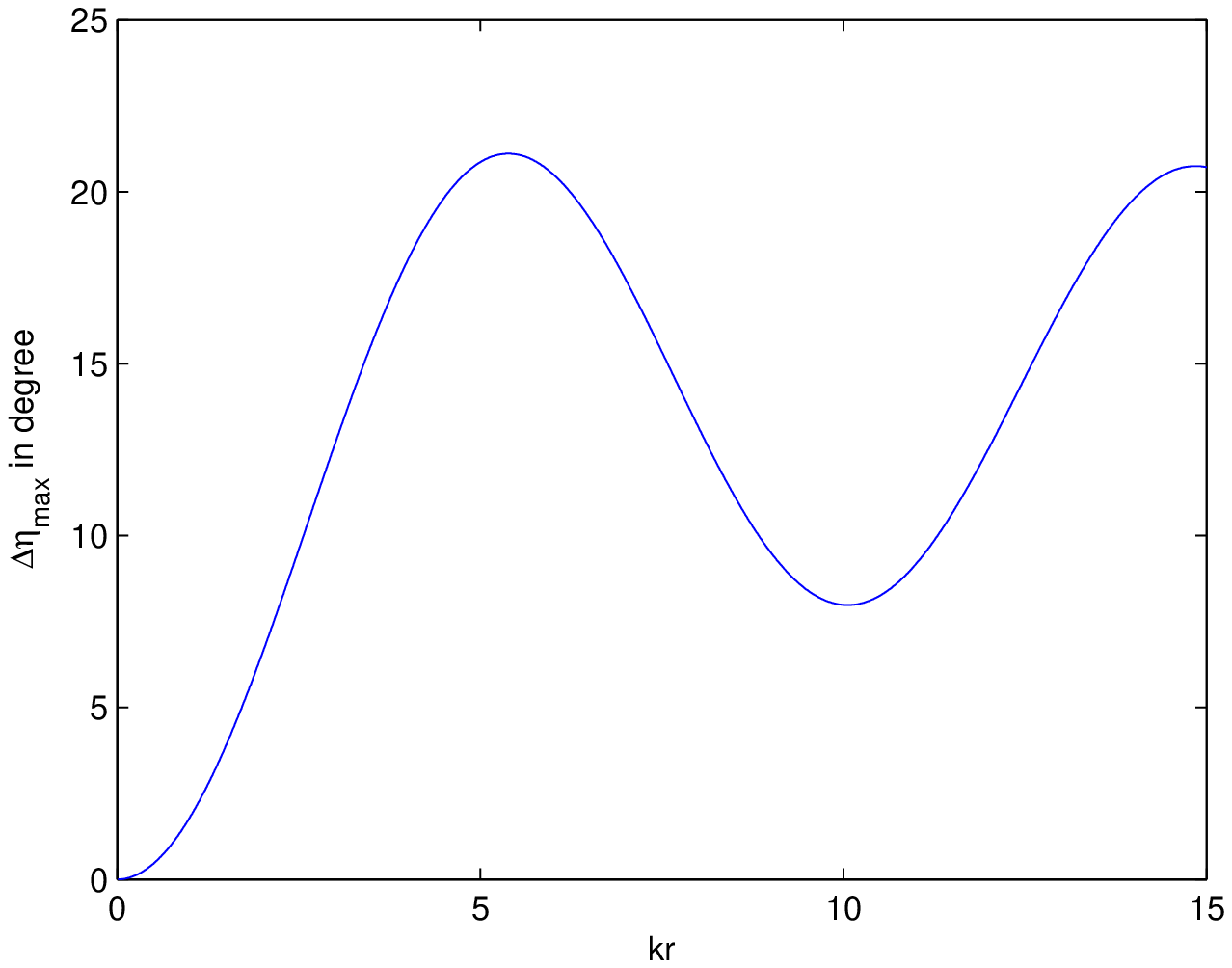}
\caption{\label{krcos2}Plot of the maximum of $\Delta\eta'$ (due to $\cos \vec{k}.\vec{r}$) vs $kr$ with $\eta_{G} = 45^{\circ}, \Psi_{0} = 0.05, \theta = 2.27$ in Eq.~(\ref{deltaetaprime}).}
\end{figure}

Radiations coming from the cosmologically distant radio galaxies are polarized \cite{Alven, *Gardner1, *Gardner2, *Burbidge}. It has been an interesting puzzle whether the observed angle($\Delta \eta$) between the polarization vector and the characteristic axis of the galaxy has some statistically significant correlation with the angular positions($\theta, \phi$) of the galaxy on the sky \cite{Birch, KY, BK, NR1, *NR2, JR, Carroll, Loredo}. By observed angle($\Delta \eta$), we mean the angle remaining after extracting the Faraday rotation. Analysis of the observed data indicates the existence of a dipolar anisotropy in the distribution of $\Delta \eta$ over the sky \cite{NR1, JR, Birch, KY, BK}. One may concult ref.~\cite{JR} for a history of puzzling observations. Surprisingly, the anisotropy axis coincides with the dipoles determined independently from the polarizations at optical frequencies, sky brightness at radio frequencies and the axes associated with CMBR multipoles (see \cite{RJ, *Kuhne} and references therein). It has been tried to explain this preferred direction as a result of the \textit{peculiar motion} of the solar system \cite{Blake, Singal}. But the peculiar velocities determined from CMBR data and from radio data are not consistent with each other \cite{Singal, Tiwari, Rubart}. This suggests the existence of an intrinsic anisotropy in the large scale with the anisotropy axis roughly aligned along the CMBR dipole \cite{Tiwari}. In recent years, single mode super-horizon perturbations have been considered \cite {ECK, EKC} to explain the observed dipolar anisotropy as the wavevector corresponding to a single mode can readily introduce a special direction in the universe. We investigate whether a single mode of scalar perturbation whose wavevector points along the direction of anisotropy can lead to the observed anisotropy in the distribution of polarization angles. \\
Since the wavelength of the radiation is much much smaller than the scale of curvature, we can use the geometrical optics approximation. A good review of geometrical optics in curved space-time is given in refs.~\cite{MTW, Brans, Paddy}. In the limit of geometrical optics, the propagation vectors form a congruence of affinely parameterized null geodesics and the polarization vector is parallel transported along the null geodesics. The expansion, rotation and shear of the image carried by the null geodesics are quantified by the \textit{optical scalars} introduced by Sachs \cite{Sachs}. The Newman-Penrose formalism \cite{NP, GHP, Chandra} provides a systematic approach to calculate the optical scalars from the \textit{spin co-efficients} without solving the equations from geometrical optics approximations.\\
In this paper, we assume the galaxies to be elliptic and provide a brief review (see \cite{KO, Obukhov, Frolov} for detail) of the rotation of the elliptic major axis with respect to the polarization vector when the image propagates through a curved spacetime. We argue that the shear and the rotation are interrelated. When shear is non-zero, the rotation depends on the initial orientation of both the major axis and polarization vector on the plane perpendicular to the propagation vector. Both rotation and shear are zero for Friedmann-Robertson-Walker (FRW) metric, which describes the homogeneous, isotropic and expanding universe. We add scalar perturbation to the FRW metric and calculate the shear and rotation for a general scalar perturbation, using linear perturbation theory. We show (in Eqs.~(\ref{deltaetafinal}) and (\ref{deltaetaprime})) that the major axis undergoes a rotation with respect to the polarization vector as the elliptic image and the polarization vector are carried through the perturbed spacetime. The rotation explicitly depends on the form of scalar perturbation. We assume a single mode of perturbation, whose wavevector is along the direction of anisotropy, to find an expression of the net rotation as a result of this mode. The rotation is independent of the frequency of the electromagnetic wave concerned, hence is not related to Faraday rotation. For perturbation with very large wavelength compared to the distance of the galaxy ($kr \ll 1$), the rotation is negligibly small. But for modes with wavelength of the order of the distance to the galaxy, the amount of rotation is appreciably large (see Fig.~\ref{krsine}, \ref{krcos1} and \ref{krcos2}). Moreover, if we plot the theoretical result due to the sine-perturbation as a function of the polar angle($\theta$), it almost looks like a sinusoid (see Fig.~\ref{thetasine}) which is similar to the claimed dipole pattern. We also show the results due to the cosine-perturbation in Fig.~\ref{thetacos1}, \ref{thetacos2}\\
Here is the outline of the paper. In Section II, we discuss the rotation and shear of an elliptic image as it propagates through a curved spacetime. In Section III, we consider FRW metric with scalar perturbation and calculate the rotation and shear of the elliptic image. In Section IV, we determine the rotation due to a single mode of scalar perturbation. Section V provides the conclusion. \\

\section{Rotation and shear of an elliptic image}

Rotation and shear of a circular image with respect to the polarization vector have been discussed in ref.~\cite{Chandra} and the case of elliptic image has been considered in refs.~\cite{Frolov, KO, Obukhov}. Here we are assuming that the major axis of the elliptic image makes an angle $\eta_{0}$ with the polarization vector, at the initial point. As the image propagates along the ray, the polarization vector is parallel transported but the major axis of the image is not. As a consequence, the angle($\eta$) between them changes. Here we will outline the derivation of the expression for the rotation of the major axis with respect to the polarization vector. We will also show that, in the presence of shear, the rotation depends on the initial configuration of the polarization vector. As a byproduct of this derivation, we will also see how the lengths of the major and minor axes of an elliptic image change as the image propagates along the ray.\\
In NP formalism, $l^{\mu}$ defines the affinely parameterized null geodesic and $\{m^{\mu}$, $m^{*\mu}\}$ span the 2-dimensional plane perpendicular to the null geodesic. The null geodesic is parameterized by an affine parameter $s$. At $s = s_{1}$, we consider an elliptic image with major axis $a$ and minor axis $b$. The points on the boundary of the ellipse with respect to the center of the ellipse are given by 
\begin{equation}
\zeta^{\mu}(s_{1},\phi) = a \cos \phi \; e^{\mu}_{1} + b \sin \phi \; e^{\mu}_{2} \label{ellipsereal}
\end{equation}
where $e^{\mu}_{1}$ and $e^{\mu}_{2}$ define the major and minor axes. As $\phi$ goes from $0$ to $2\pi$, $\zeta^{\mu}(s_{1}, \phi)$ describes the boundary of the ellipse at $s = s_{1}$. For better physical understanding, we construct two real orthonormal basis vectors from $m^{\mu}$ and $m^{*\mu}$.
\begin{eqnarray}
E^{\mu} &&= \frac{1}{\sqrt{2}} (m^{\mu} + m^{*\mu}) \nonumber \\
F^{\mu} &&= \frac{1}{i\sqrt{2}} (m^{\mu} - m^{*\mu}) \label{basis}
\end{eqnarray}
For a generalized description, we want the major axis($e^{\mu}_{1}$) to make an angle $\eta_{0}$ with $E^{\mu}$ which is equivalent to the polarization vector i.e. the real part of $m^{\mu}$.
\begin{eqnarray}
e^{\mu}_{1} &&= \cos \eta_{0} \; E^{\mu} + \sin \eta_{0} \; F^{\mu} \nonumber \\
e^{\mu}_{2} &&= - \sin \eta_{0} \; E^{\mu} + \cos \eta_{0} \; F^{\mu} \label{ellaxes}
\end{eqnarray}
Writing $\zeta^{\mu}(s_{1},\phi)$ in terms of $m^{\mu}$ and $m^{*\mu}$, we get the following.
\begin{equation} 
\zeta^{\mu}(s_{1}, \phi) = \zeta(\phi) m^{*\mu} + \zeta^{*}(\phi)m^{\mu} \label{ellipsevec}
\end{equation}
where
\begin{equation}
\zeta(\phi) = \frac{1}{\sqrt{2}} \; e^{i\eta_{0}} \Big ( \frac{a + b}{2} e^{i\phi} + \frac{a-b}{2} e^{-i\phi} \Big ) \label{ellipsescalar}
\end{equation}
At the neighbouring point, $s_{2} = s_{1} + \delta s$, the boundary of the image is given by (see ref.~\cite{Frolov} for proof):
\begin{equation}
\zeta^{\mu}(s_{2},\phi) = \zeta'(\phi) m^{*\mu} + \zeta'^{*}(\phi) m^{\mu} \label{ellipsevecs2}
\end{equation}
where
\begin{equation}
\zeta'(\phi) = \zeta(\phi) - \delta s \big [ \rho \; \zeta(\phi) + \sigma \; \zeta^{*}(\phi) \big ] \label{ellipsescalars2}
\end{equation}
where $\rho$ and $\sigma$ are the spin-coefficients (see Eqs.~(\ref{rho}) and (\ref{sigma})). Now we will study the change from $\zeta(\phi)$ to $\zeta'(\phi)$. We can write $\zeta(\phi)$ given by Eq.~(\ref{ellipsescalar}) as:
\begin{equation}
\zeta(\phi) = \frac{1}{\sqrt{2}} \; e^{i\eta_{0}} \; \big ( p_{1} e^{i\phi} + p_{2} e^{-i\phi} \big ) \label{ellipsescalarp}
\end{equation}
where $p_{1} = \frac{(a+b)}{2}$ and $p_{2} = \frac{(a-b)}{2}$. We can get the lengths of major and minor axes of the elliptic image at $s = s_{1}$ by linear combination of $p_{1}$ and $p_{2}$. Similarly, at $s = s_{2}$, $\zeta'(\phi)$ can be written as:
\begin{equation}
\zeta'(\phi) = \frac{1}{\sqrt{2}} \; e^{i\eta_{0}} \; \big ( P_{1} e^{i\phi} + P_{2} e^{-i\phi} \big ) \label{ellipsescalarps2}
\end{equation}
where $P_{1}$ and $P_{2}$ are given by:
\begin{subequations}
\begin{eqnarray}
P_{1} &&= p_{1} - \delta s (\rho \; p_{1} + e^{-2i\eta_{0}} \; \sigma \; p_{2}), \label{P1}\\
P_{2} &&= p_{2} - \delta s (\rho \; p_{2} + e^{-2i\eta_{0}} \; \sigma \; p_{1}) \label{P2}
\end{eqnarray}
\end{subequations}
Here, $P_{1}$ and $P_{2}$ are complex, in general, due to the presence of $\rho$ and $\sigma$. We separate the magnitudes and phases in $P_{1}$ and $P_{2}$ by writing $P_{1} = |P_{1}| e^{i\chi_{1}}$ and $P_{2} = |P_{2}| e^{i\chi_{2}}$. $|P_{1,2}|$ and $\chi_{1,2}$ are given by:
\begin{subequations}
\begin{eqnarray}
|P_{1}| &&= p_{1} \Big [ 1 - \delta s \Big \{ \text{Re} \; \rho + \frac{p_{2}}{p_{1}} \big ( \cos 2\eta_{0} \; \text{Re} \; \sigma + \sin 2\eta_{0} \; \text{Im} \; \sigma \big )\Big \} \Big ], \label{P1mod}\\
\chi_{1} &&= - \delta s \Big [ \text{Im}\; \rho + \frac{p_{2}}{p_{1}} \big ( \cos 2\eta_{0} \; \text{Im} \; \sigma - \sin 2\eta_{0} \; \text{Re} \; \sigma \big ) \Big ], \label{chi1}\\
|P_{2}| &&= p_{2} \Big [ 1 - \delta s \Big \{ \text{Re} \; \rho + \frac{p_{1}}{p_{2}} \big ( \cos 2\eta_{0} \; \text{Re} \; \sigma + \sin 2\eta_{0} \; \text{Im} \; \sigma \big )\Big \} \Big ], \label{P2mod}\\
\chi_{2} &&= - \delta s \Big [ \text{Im}\; \rho + \frac{p_{1}}{p_{2}} \big ( \cos 2\eta_{0} \; \text{Im} \; \sigma - \sin 2\eta_{0} \; \text{Re} \; \sigma \big ) \Big ], \label{chi2}
\end{eqnarray}
\end{subequations}
From $|P_{1}|$ and $|P_{2}|$, we can calculate the lengths of major and minor axes of the elliptic image at $s = s_{2}$. The phase factors associated with $P_{1}$ and $P_{2}$ can be used to extract the information about the rotation of the major and minor axes with respect to the polarization vector. If the changes in lengths of major and minor axes are $\delta a$ and $\delta b$ respectively, then
\begin{subequations}
\begin{eqnarray}
\frac{\delta a}{a} &&= - (\text{Re}\;\rho + \cos 2\eta_{0} \; \text{Re} \; \sigma + \sin 2\eta_{0} \; \text{Im} \; \sigma)\delta s, \label{deltaa} \\
\frac{\delta b}{b} &&= - (\text{Re}\;\rho - \cos 2\eta_{0} \; \text{Re} \; \sigma - \sin 2\eta_{0} \; \text{Im} \; \sigma)\delta s \label{deltab}
\end{eqnarray}
\end{subequations}
One should note that $\frac{\delta a}{a}$ and $\frac{\delta b}{b}$ are not equal due to the presence of $\sigma$ which implies a non-zero shear. In the absence of $\sigma$, it would have been a uniform expansion or contraction of the image. To understand the rotation of the images, we write $\zeta'(\phi)$ as:
\begin{equation}
\zeta'(\phi) = \frac{1}{\sqrt{2}} \exp \Big \{ i \Big ( \eta_{0} + \frac{\chi_{1} + \chi_{2}}{2} \Big ) \Big \}. \Big [ |P_{1}| \exp \Big \{ i \Big ( \phi + \frac{\chi_{1} - \chi_{2}}{2} \Big ) \Big \} +  |P_{2}| \exp \Big \{ -i \Big ( \phi + \frac{\chi_{1} - \chi_{2}}{2} \Big ) \Big \} \Big ] \label{zetaprime}
\end{equation}
The modification from $\phi$ to $\{ \phi + (\chi_{1} - \chi_{2})/2 \}$ is not measurable as it only changes the starting point of $\phi$. But, if we look at Eq.~(\ref{ellipsevecs2}) for $\zeta^{\mu}(s_{2},\phi)$, we see that the emergence of the overall phase factor in $\zeta'(\phi)$ is equivalent to a rotation of $\{ e_{1}^{\mu}, e_{2}^{\mu} \}$ into $\{ e'^{\mu}_{1}, e'^{\mu}_{2} \}$.
\begin{subequations}
\begin{eqnarray}
e'^{\mu}_{1} &&= \cos \Big ( \frac{\chi_{1} + \chi_{2}}{2} \Big ) e_{1}^{\mu} + \sin \Big ( \frac{\chi_{1} + \chi_{2}}{2} \Big ) e_{2}^{\mu}, \label{e1prime} \\
e'^{\mu}_{2} &&= - \sin \Big ( \frac{\chi_{1} + \chi_{2}}{2} \Big ) e_{1}^{\mu} + \cos \Big ( \frac{\chi_{1} + \chi_{2}}{2} \Big ) e_{2}^{\mu} \label{e2prime}
\end{eqnarray}
\end{subequations}
In terms of $\{ e'^{\mu}_{1} , e'^{\mu}_{2} \}$, $\zeta^{\mu}(s_{2},\phi)$ can be written as:
\begin{equation}
\zeta^{\mu}(s_{2},\phi) = \big ( |P_{1}| + |P_{2}| \big ) \cos \Big ( \phi + \frac{\chi_{1} - \chi_{2}}{2} \Big ) e'^{\mu}_{1} + \big ( |P_{1}| - |P_{2}| \big ) \sin \Big ( \phi + \frac{\chi_{1} -  \chi_{2}}{2} \Big ) e'^{\mu}_{2} \label{zetavec}
\end{equation} 
Therefore, at $s = s_{2}$, $\{ e'^{\mu}_{1} , e'^{\mu}_{2} \}$ define the major and minor axes of the elliptic image and the major axis ($e'^{\mu}_{1}$) makes an angle of $\big ( \eta_{0} + \frac{\chi_{1} + \chi_{2}}{2} \big )$ with the polarization vector ($E^{\mu}$). Since $m^{\mu}$ is parallel transported along the null geodesic, it does not rotate. The rotation, $\delta \eta$, of the principal axes with respect to the polarization vector is given by
\begin{equation}
\delta \eta = \frac{1}{2} \big ( \chi_{1} + \chi_{2} \big ) = - \delta s \Big \{ \text{Im}\;\rho + \frac{a^{2} + b^{2}}{a^{2} - b^{2}} \big ( \cos 2\eta_{0} \; \text{Im}\;\sigma - \sin 2\eta_{0} \; \text{Re} \; \sigma \big ) \Big \} \label{deltaeta}
\end{equation}
One should note that, due to the presence of the shear ($\sigma$), the net rotation of major and minor axes depends on the initial angle $(\eta_{0})$ between the polarization vector and the major axis. Before we conclude this section, we would like to discuss the consequences of the transformation: $m^{\mu} \rightarrow \tilde{m}^{\mu} = e^{-i\psi} m^{\mu}$. Before this transformation, the major axis $(e^{\mu}_{1})$ makes angle $\eta_{0}$ with respect to the polarization vector $(E^{\mu})$ in the anticlockwise direction. This transformation implies an anticlockwise rotation of $\{ E^{\mu} , F^{\mu} \}$ by an angle $\psi$. Since ${m}^{\mu} = e^{i\psi} \tilde{m}^{\mu}$, writing $\zeta^{\mu}(s_{1},\phi)$ (in Eq.~(\ref{ellipsevec})) in terms of $\{\tilde{m}^{\mu}, \tilde{m}^{*\mu}\}$, one would get:
\begin{equation} 
\zeta^{\mu}(s_{1}, \phi) = \tilde{\zeta}(\phi) \tilde{m}^{*\mu} + \tilde{\zeta}^{*}(\phi)\tilde{m}^{\mu} \label{ellipsevectilde}
\end{equation}
where
\begin{equation}
\tilde{\zeta}(\phi) = \frac{1}{\sqrt{2}} \; e^{i(\eta_{0} - \psi)} \Big ( \frac{a + b}{2} e^{i\phi} + \frac{a-b}{2} e^{-i\phi} \Big ) \label{ellipsescalartilde}
\end{equation}
This implies that the major axis makes an angle of $(\eta_{0} - \psi)$ with the transformed polarization vector $(\tilde{E}^{\mu})$. After this transformation, $\delta \eta$ of Eq.~(\ref{deltaeta}) will be modified, as $\eta_{0}$ will be replaced by $(\eta_{0} - \psi)$.

\section{Null geodesics in perturbed FRW metric}

In this section, we will consider large scale scalar perturbations to the FRW metric. Any scalar perturbation can be written as a superposition of infinite number of Fourier modes i.e. $ \Psi (t, \vec{r}) = \int \frac{d^{3}k}{(2\pi)^{3}} \; e^{i\vec{k}.\vec{r}} \; \tilde{\Psi} (t, \vec{k}) $. Our goal is to study the effect of a single mode of scalar perturbation on the net rotation of the major axis of an elliptic image with respect to the polarization vector. The co-ordinate system is chosen such that the wave vector($\vec{k}$) corresponding to the specific mode of the perturbation is along the z-axis i.e. $\vec{k} = k \hat{z}$. Then  the scalar perturbation($\Psi$) will have the following form.
\begin{equation}
\Psi = \Psi_{0}(t,k) \sin (k r \cos \theta) + \Psi'_{0}(t,k) \cos (k r \cos \theta) \label{Psi}
\end{equation}
where $\theta$ is the angle between the position vector($\vec{r}$) and the z-axis. Since the perturbation($\Psi$) does not depend upon the azimuthal angle($\phi$), the geometry will have azimuthal symmetry. Motivated by this fact, we will consider the scalar perturbations which depend on $t, r, \theta$ but not on $\phi$. \\
In the Newtonian gauge \cite{Dodelson}, we introduce two large scale scalar perturbations, $\Psi$ and $\Phi$, to the FRW metric. The line element in the spherical polar co-ordinates takes the following form.
\begin{equation}
ds^{2} = (1 + 2 \Psi ) dt^{2} - a^{2}(t) (1 - 2 \Phi)  (dr^{2} + r^{2} d\theta^{2} + r^{2} \sin^{2} \theta d\phi^{2} ) \label{dsperturbed}
\end{equation}
where $\Psi = \Psi (t,r,\theta)$ and $\Phi = \Phi(t,r,\theta)$. Our chosen null vectors are the following.
\begin{subequations}
\begin{eqnarray}
l^{\mu} &&\equiv (\frac{1}{a} + \epsilon_{t} ) \hat{t} + \frac{1}{a^{2}} (1 + a \epsilon_{t} + \Psi + \Phi) \hat{r} + \epsilon_{\theta} \hat{\theta}, \label{lmunew} \\
n^{\mu} &&\equiv \frac{a^{2}}{2} (\frac{1}{a} - \epsilon_{t} - \frac{2 \Psi}{a}) \hat{t} - \frac{1}{2} (1 - a \epsilon_{t} - \Psi + \Phi) \hat{r} - \frac{a^{2}}{2} \epsilon_{\theta} \hat{\theta}, \label{nmunew} \\
m^{\mu} &&\equiv - \frac{i}{\sqrt{2}} a r \epsilon_{\theta} \hat{r} + \frac{i}{\sqrt{2} a r} (1 + \Phi) \hat{\theta} + \frac{1}{\sqrt{2} a r \sin \theta}(1 + \Phi) \hat{\phi} \label{mmunew}
\end{eqnarray}
\end{subequations}
$\epsilon_{t}$ and $\epsilon_{\theta}$ are defined in Eqs.~(\ref{epsilont})-(\ref{epsilontheta}). Here, $m^{\mu}$ is defined upto a constant phase factor. Now we calculate the relevant spin co-efficients.
\begin{subequations}
\begin{eqnarray}
\kappa &&= 0, \label{kappa0} \\
\epsilon &&= 0, \label{epsilon0}\\
\pi &&= \frac{-i}{\sqrt{2} a r } \partial_{\theta} \Psi \label{piexp}
\end{eqnarray}
\end{subequations}
Eqs.~(\ref{kappa0})-(\ref{epsilon0}) imply that $l^{\mu}$ forms a congruence of null geodesics which are affinely parameterized. Since $\pi \neq 0$, $m^{\mu}$ and $m^{*\mu}$ are not parallel transported along the null geodesics (see Eqs.~(\ref{kappa})-(\ref{epsilon})). The other two spin co-efficients are the following.
\begin{subequations}
\begin{eqnarray}
\rho =&& -H\Big  (\frac{1}{a} + \epsilon_{t}\Big ) - \frac{1}{a^{2} r} (1 + \Psi + \Phi) - \frac{\epsilon_{t}}{a r} \nonumber \\
&&- \frac{1}{2} \cot \theta \;\; \epsilon_{\theta} - \frac{1}{2} \partial_{\theta} \epsilon_{\theta} + \frac{1}{a} \partial_{t} \Phi + \frac{1}{a^{2}} \partial_{r} \Phi, \label{rhoexp}\\
\sigma =&& \frac{1}{2} (\partial_{\theta} \epsilon_{\theta} - \cot \theta \epsilon_{\theta}) \label{sigmaexp}
\end{eqnarray}
\end{subequations}
A non-zero $\sigma$ implies that the geometry is not free from shear. To make $\pi = 0$, we need to apply a rotation of class I (see Eqs.~(\ref{vec1})-(\ref{spin1})). Under this rotation, the null vectors and the spin coefficients change according to Eqs.~(\ref{vec1})-(\ref{spin1}). Since before the transformation, $\kappa$ and $\epsilon$ were zero, they remain zero and the expressions for $\rho$ and $\sigma$ do not change. But the expression for $\pi$ changes and $z$ is chosen such that the new $\pi$ becomes zero.
\begin{equation}
\pi \rightarrow \frac{-i}{\sqrt{2} a r} \partial_{\theta} \Psi + l^{\mu} \partial_{\mu} z^{*} = 0 \label{pi0}
\end{equation}
We do not need to solve for $z$ as it does not appear in the expressions of $\rho$ and $\sigma$. The rotation $\delta\eta$ is given by Eq.~(\ref{deltaeta}) where $\rho$ and $\sigma$ are given by Eqs.~(\ref{rhoexp})-(\ref{sigmaexp}) and $\eta_{0}$ is the angle between the polarization vector and the major axis at the location of galaxy. As explained below, the polarization vector is in the direction of $\hat{\phi}$, hence $\eta_{0}$ is the angle between the major axis and $\hat{\phi}$, at the location of galaxy.\\
The polarization vector is the real part of $m^{\mu}$. In three dimensions, the propagation vector($\vec{l}$) is the spatial part of $l^{\mu}$ and the polarization vector($\vec{m}$) is the spatial part of $\text{Re} \; m^{\mu}$. The proper set of null vectors are the ones obtained after applying a rotation of class I (see Eqs.~(\ref{vec1})-(\ref{spin1})) to the null vectors of Eqs.~(\ref{lmunew})-(\ref{mmunew}). We did not solve the complex parameter($z$) of this rotation, because it does not appear in the expression of the net rotation. If we determine the polarization vector($\vec{m}$), $z$ does appear in $\vec{m}$. But the component of $\vec{m}$ which is perpendicular to the propagation vector($\vec{l}$), is independent of $z$, and that component turns out to be the real part of Eq.~(\ref{mmunew}).
\begin{equation}
\vec{m}_{\perp} = \frac{1}{\sqrt{2} a r \sin \theta} (1  + \Phi) \hat{\phi} \label{m3vec}
\end{equation}
The light is propagating along the direction of $(- \hat{r})$. In three dimensions, the plane perpendicular to $\hat{r}$ consists of two perpendicular vectors, $\hat{\theta}$ and $\hat{\phi}$. The most general orientation of $\vec{m}_{\perp}$ would be a combination of $\hat{\theta}$ and $\hat{\phi}$. After the transformation, $m^{\mu} \rightarrow e^{-i\psi} m^{\mu}$, $\vec{m}_{\perp}$ takes the following form in the $\hat{\theta}$-$\hat{\phi}$ plane.
\begin{equation}
\vec{m}_{\perp} = \frac{(1 + \Phi)}{\sqrt{2} a} \Big ( \cos \psi \; \frac{\hat{\phi}}{r \; \sin \theta} + \sin \psi \; \frac{\hat{\theta}}{r} \Big ) \label{m3vecnew}
\end{equation}
Therefore, $\psi$ is the angle between the polarization vector and $\hat{\phi}$ at the location of the galaxy and $(\eta_{0} - \psi)$ is the angle between the major axis and the polarization vector. Hence $\delta \eta$ is modified to the following expression.
\begin{equation}
\delta \eta = \delta s \frac{\sin (2\eta_{0} - 2\psi)}{2} (\partial_{\theta} \epsilon_{\theta} - \cot \theta \; \epsilon_{\theta}) \label{deltaetaexp}
\end{equation}
where $\delta s$ is the change in affine parameter. We have assumed the size of the minor axis to be very small compared to that of the major axis i.e. $b \ll a$ in Eq.~(\ref{deltaeta}). Writing $\epsilon_{\theta} = \frac{\tilde{\epsilon}}{a^{2}}$, and assuming two scalar perturbations to be equal, $\Psi = \Phi$ (which is the case when anisotropic stress is zero \cite{Dodelson}), we have the following equation (see Eq.~(\ref{epsilonthetaeqn})).
\begin{equation}
\partial_{r} (r^{2} \tilde{\epsilon}) = - 2 \; \partial_{\theta} \Psi \label{etildepsi}
\end{equation}
Then the rotation becomes
\begin{equation}
\delta \eta = \delta s \frac{\sin (2\eta_{G})}{2a^{2}} (\partial_{\theta} \tilde{\epsilon} - \cot \theta \; \tilde{\epsilon}) \label{deltaetaetilde}
\end{equation}
where $\eta_{G} = \eta_{0} - \psi$ is the angle between the major axis and the polarization vector at the location of the galaxy. So when the scalar perturbation, $\Psi$, is given, we have to solve for $\tilde{\epsilon}$ using Eq.~(\ref{etildepsi}) and using that, we can determine $\delta \eta$. \\
To get the net rotation as the light travels from the source to the observer, we have to integrate Eq.~(\ref{deltaetaetilde}) for a radial null geodesic. Since $\dot{r} = \frac{d r }{d s }$ was taken to be positive, $r$ increases with $s$. We are assuming that the affine parameter decreases as the light propagates from the galaxy to the observer. This allows us to assume the observer to be at the origin ($r=0$) and the source to be at a distance $r$ from the observer. Since $\dot{r} = \frac{1}{a^{2}} + \frac{\epsilon_{t}}{a} + \frac{2 \Psi}{a^{2}}$, we can write $d s $ as the following in the first order.
\begin{equation}
d s = \frac{ds}{dr} dr = a^{2} (1 - a \epsilon_{t} - 2 \Psi) dr \label{ds}
\end{equation}
So, in first order, $d \eta$ becomes
\begin{equation}
d\eta = dr \frac{\sin (2\eta_{G})}{2} (\partial_{\theta} \tilde{\epsilon} - \cot \theta \tilde{\epsilon}) \label{deta}
\end{equation}
Now we are integrating this equation assuming the angle($\eta$) to be $\eta_{G}$ at the location of the galaxy($r$) and $\eta = \eta_{G} + \Delta \eta$ at the location of the observer($r=0$).
\begin{equation}
\Delta \eta = - \frac{\sin (2\eta_{G})}{2} \int_{0}^{r} dr (\partial_{\theta} \tilde{\epsilon} - \frac{\cos \theta}{\sin \theta} \tilde{\epsilon}) \label{deltaetaint}
\end{equation}
This is a generic result valid for any scalar perturbation. To summarize, $\tilde{\epsilon}$ can be found by solving Eq.~(\ref{etildepsi}), $\eta_{G}$ is the angle between the major axis and the polarization vector at the location of the galaxy.

\section{Net rotation of polarization vector due to a single mode of scalar perturbation}

In this section, we consider a single mode of scalar perturbation. The wavevector($\vec{k}$) corresponding to the mode is assumed to point toward the $z$-axis i.e. $\vec{k} = k \hat{z}$. First, we consider the effect of the \textit{sine}-perturbation.
\begin{equation}
\Psi = \Psi_{0} (t, k) \sin (\vec{k}.\vec{r}) = \Psi_{0} \sin (k r \cos \theta) \label{Psiexp}
\end{equation}
Now we solve Eq.~(\ref{etildepsi}) to solve for $\tilde{\epsilon}$. The most general solution of $\tilde{\epsilon}$ is
\begin{equation}
\tilde{\epsilon} = 2 \Psi_{0} \; \frac{\sin \theta}{\cos \theta} \; \frac{1}{r} \Big ( \sin (k r \cos \theta)  + \frac{\cos (k r \cos \theta)}{k r \cos \theta}\Big ) + \frac{f(\theta)}{r^{2}} \label{etildesol}
\end{equation}
where $f(\theta)$ appears as a constant of integration. We fix $f(\theta)$ such that $\tilde{\epsilon}$ does not diverge at $r \rightarrow 0$. 
\begin{equation}
f(\theta) = - 2 \Psi_{0} \frac{\sin \theta}{k \cos ^{2} \theta} \label{ftheta}
\end{equation}
Then $\tilde{\epsilon}$ has the following form.
\begin{equation}
\tilde{\epsilon} = 2 \Psi_{0} \; \frac{\sin \theta}{\cos \theta} \; \frac{1}{r} \Big[ \sin (k r \cos \theta) - \frac{2}{k r \cos \theta} \sin ^{2} \Big( \frac{k r}{2} \cos \theta \Big ) \Big ] \label{etildeexp}
\end{equation}
Now the calculation of $\Delta \eta$ from Eq.~(\ref{deltaetaint}) is straightforward. 
\begin{equation}
\Delta \eta = - \sin (2\eta_{G}) \; \Psi_{0} \; \frac{\sin ^{2}\theta}{\cos ^{2} \theta} \; \Big [ \frac{4}{k r \cos \theta} \sin ^{2} \Big (\frac{k r}{2} \cos \theta \Big ) - \sin (k r \cos \theta)\Big ] \label{deltaetafinal}
\end{equation}
We emphasize that there is no singularity in $\Delta \eta$. For a given value of $k r$, if we plot $\Delta \eta$ vs $\theta$, it looks almost like a sinusoid (see Fig.~\ref{thetasine}) which qualitatively agrees with the previous claim of dipole signature \cite{KY, BK, JR}. The magnitude of the maximum of $\Delta \eta$ increases with $k r$ (see Fig.~\ref{krsine}). If the wavelength ($\lambda = \frac{2 \pi}{k}$) of the mode of perturbation is very large compared to the distance($r$) of the galaxy ($k r \ll 1$), $\Delta \eta$ tends to be zero. But for the modes with wavelength comparable to the distance of the galaxy, $\Delta \eta$ is appreciably large e.g. with $k r = 6, \; \eta_{G}= 45^{\circ}, \; \Psi_{0} = 0.05$; $\Delta \eta_{max} = 11.72^{\circ}$ and with $k r = 12, \; \eta_{G}= 45^{\circ}, \; \Psi_{0} = 0.05$; $\Delta \eta_{max} = 53.19^{\circ}$. \\
Another component of the Fourier mode is $\Psi'_{0} (t, k) \cos (\vec{k}.\vec{r})$. From Eq.~(\ref{etildepsi}), we get the most general solution of $\tilde{\epsilon}$. 
\begin{equation}
\tilde{\epsilon} = 2 \Psi_{0} \; \frac{\sin \theta}{\cos \theta} \; \frac{1}{r} \Big ( \cos (k r \cos \theta)  - \frac{\sin (k r \cos \theta)}{k r \cos \theta}\Big ) \label{etildecos}
\end{equation}
Here the constant of integration has been fixed to zero so that $\tilde{\epsilon}$ remains finite as $r \rightarrow 0$. Now one can calculate $\Delta \eta$ for cosine-perturbation from Eq.~(\ref{deltaetaint}).
\begin{equation}
\Delta \eta' = - \sin (2\eta_{G}) \; \Psi'_{0} \; \frac{\sin ^{2}\theta}{\cos ^{2} \theta} \; \Big [ \frac{2}{\cos \theta} \sin ^{2} \Big (\frac{k r}{2} \cos \theta \Big ) + \frac{2 \sin (k r \cos \theta)}{k r \cos \theta} - 2 \Big ] \label{deltaetaprime}
\end{equation}
$\Delta \eta'$ diverges as $\theta \rightarrow \frac{\pi}{2}$. We plot $\Delta \eta'$ vs $\theta$ for two regions, $\theta < \frac{\pi}{2}$ (see Fig.~\ref{thetacos1}) and $\theta > \frac{\pi}{2}$ (see Fig.~\ref{thetacos2}). With $k r = 12, \; \eta_{G}= 45^{\circ}, \; \Psi_{0} = 0.05$; $|\Delta \eta'|$ has a local maximum of $18.13^{\circ}$ at $\theta = 1.07$ (for $\theta < \frac{\pi}{2}$) and $12.54^{\circ}$ at $\theta = 2.27$ (for $\theta > \frac{\pi}{2}$). $\Delta \eta'$ vs $k r$ has been plotted for $\theta = 1.07$ (see Fig.~\ref{krcos1}) and $\theta = 2.27$ (see Fig.~\ref{krcos2}). Unlike $\Delta \eta_{max}$, $\Delta \eta'_{max}$ is a periodic function of $k r$. For $\theta = 1.07$, $\Delta \eta'_{max} \approx 20^{\circ}$ and for $\theta = 2.27$, $\Delta \eta'_{max} \approx 21^{\circ}$ and $\Delta \eta'_{min} \approx 8^{\circ}$ (excluding $\Delta \eta' = 0$ for $k r = 0$). Since we are doing linear perturbation theory, $\Delta \eta$ due to the perturbation, $\Psi = \Psi_{0} \sin(\vec{k}.\vec{r}) + \Psi'_{0} \cos(\vec{k}.\vec{r})$, can simply be written as the sum of two $\Delta \eta$'s: $\Delta \eta_{total} = \Delta \eta + \Delta \eta'$.\\
Eqs.~(\ref{deltaetafinal}) and (\ref{deltaetaprime}) are the main results of the paper. As we can see, the net rotation does depend upon the angular position($\theta$) of the galaxy and the distance($r$) of the galaxy as well as the the orientation of the major axis and the polarization vector. This rotation is independent of the frequency of the light emitted from the galaxy unlike Faraday rotation. This rotation is a result of the large scale anisotropic scalar perturbation to the homogeneous, isotropic universe. Eqs.~(\ref{deltaetaint}) and (\ref{etildepsi}) provides the net rotation ($\Delta \eta$) due to a scalar perturbation ($\Psi$). Then we consider a single Fourier mode of the scalar perturbation which consists of $\sin (\vec{k}.\vec{r})$ and $\cos (\vec{k}.\vec{r})$. We have described the effects of \textit{sine-} and \textit{cosine-}perturbations. Since the expression of the rotation is linear in scalar perturbation and a scalar pertubation is nothing but linear combination of various Fourier modes, it is easy to calculate the effect of any perturbation using Eqs.~(\ref{deltaetafinal}) and (\ref{deltaetaprime}).

\section{Conclusion}

We have shown how a large scale scalar perturbation ($\Psi$) to the FRW metric can lead to the residual rotation ($\Delta \eta$) of an image with respect to the polarization vector as the radiation traverses large distance through the perturbed spacetime. We have derived an expression for the rotation of the major axis of an elliptic image as the result of a single mode of scalar perturbation. For each galaxy at an angular position ($\theta, \phi$), $\eta_{G}$ is the angle between the polarization vector and the major axis at the location of the galaxy. In general, this would vary from galaxy to galaxy. So $\eta_{G}$ is an unknown functions of ($\theta, \phi$). To get a rough idea of how the rotation depends on the angular co-ordinates, we have fixed $\eta_{G} = 45^{\circ}$ and plotted the rotation($\Delta \eta$) with $\theta$ for a fixed value of $kr$. Coincidentally, the result roughly demonstrates a  dipole signature under a fixed value of $\eta_{G}$. The dipole signature might go away if $\eta_{G}$ varies rapidly for different galaxies. But the rotation as a result of scalar perturbation will always retain some correlation with the angular positions. We have also shown that, for modes with wavelength much larger than the distance to the galaxy, the rotation is negligibly small; but for wavelength of the order of the distance to the galaxy, the magnitude of the rotation is appreciably large. \\
In summary, we provide a physical explanation behind the angular dependence of residual angle between the polarization and the major axis of galactic image after subtracting Faraday rotation, for the case of cosmologically distant radio galaxies. In future, it can be checked whether the observed data is consistent with the theoretical prediction.\\

\appendix
\section{Newman-Penrose Formalism in the Context of Geometrical Optics}

\subsection{Geometrical optics in curved spacetime}

Let us assume that an electromagnetic wave follows a curve $x^{\mu}(s)$, where $s$ is the affine parameter. The vector potential is written as $A^{\mu}$, which is a complex and spacelike vector. Propagation vector, $k^{\mu}$, is defined as the tangent to the curve followed by the electromagnetic wave. Polarization vector, $f^{\mu}$, is defined \cite{MTW} as the normalized vector potential.
\begin{eqnarray}
k^{\mu} &=& \frac{dx^{\mu}}{ds}, \label{kdef} \\
f^{\mu} &=& \frac{A^{\mu}}{\sqrt{-A^{\mu}A_{\mu}^{*}}} \label{fdef}
\end{eqnarray}
The negative sign is there within the square root because we are working with $(+, -, -, -)$ signature. Under the geometrical optics approximation, propagation and polarization vectors satisfy the following set of equations \cite{MTW, Paddy}.
\begin{subequations}
 \label{allequations}
 \begin{eqnarray}
 k^{\mu}k_{\mu} &=& 0, \label{gopa} \\
 f^{\mu}f_{\mu}^{*} &=& -1, \label{gopb}\\
 k^{\mu}f_{\mu} &=& 0, \label{gopc}\\
 k^{\nu}\nabla_{\nu}k^{\mu} &=& 0, \label{gopd}\\
 k^{\nu}\nabla_{\nu}f^{\mu} &=& 0 \label{gope}
\end{eqnarray}
\end{subequations}
Here $\nabla_{\mu}$ implies covariant derivative with respect to $x^{\mu}$. From Eq.~(\ref{gopb}), we see that the polarization vector($f^{\mu}$) is normalized by definition. Eq.~(\ref{gopa}) implies that the propagation vector($k^{\mu}$) is a null vector. Eq.~(\ref{gopc}) implies that the polarization vector is perpendicular to the propagation vector. Eq.~(\ref{gopd}) implies that the electromagnetic wave follows a null geodesic which is affinely parameterized. Finally, Eq.~(\ref{gope}) implies that the polarization vector is parallel transported along the null geodesic.

\subsection{Newman-Penrose formalism}

Here we will give a brief outline (see refs.~\cite{NP, GHP, Chandra} for detail) of Newman-Penrose (NP henceforth) formalism and its relevance in the context of our problem. We are following the conventions and notations of ref.~\cite{Chandra}. For a given metric, one needs to find out a set of four null vectors ($l^{\mu}, n^{\mu}, m^{\mu}, m^{*\mu}$) where $l^{\mu}$, $n^{\mu}$ are real null vectors and $m^{\mu}$, $m^{*\mu}$ are a pair of complex conjugate null vectors. They need to be constructed in a way that they satisfy the following conditions.
\begin{subequations}
\begin{eqnarray}
l^{\mu}l_{\mu} = n^{\mu}n_{\mu} &=& m^{\mu}m_{\mu} = m^{*\mu}m^{*}_{\mu} = 0, \label{null} \\
l^{\mu}m_{\mu} = l^{\mu}m^{*}_{\mu} &=& n^{\mu}m_{\mu} = n^{\mu}m^{*}_{\mu} = 0, \label{normal} \\
l^{\mu}n_{\mu} &=& 1, \;\;\; m^{\mu}m^{*}_{\mu} = -1 \label{normalized}
\end{eqnarray}
\end{subequations}
If we compare $l^{\mu}$ with the propagation vector($k^{\mu}$) and $m^{\mu}$ with the polarization vector($f^{\mu}$), we see that, by construction, $l^{\mu}$ and $m^{\mu}$ satisfy Eqs.~(\ref{gopa})-(\ref{gopc}). Now we will state the conditions for $l^{\mu}$ and $m^{\mu}$ to satisfy Eqs.~(\ref{gopd})-(\ref{gope}). \\
1. $l^{\mu}$ forms a congruence of null geodesics if 
\begin{equation}
\kappa = m^{\mu}l^{\nu}\nabla_{\nu}l_{\mu} = 0 \label{kappa}
\end{equation}
2. Provided $\kappa = 0$, the geodesics will be affinely parameterized if 
\begin{equation}
\epsilon = n^{\mu}l^{\nu}\nabla_{\nu}l_{\mu} + m^{\mu}l^{\nu}\nabla_{\nu}m^{*}_{\mu} = 0 \label{epsilon}
\end{equation}
3. Given $\kappa = \epsilon = 0$, $m^{\mu}$ and $m^{*\mu}$ will be parallel transported along $l^{\mu}$ if 
\begin{equation}
\pi = n^{\mu}l^{\nu}\nabla_{\nu}m^{*}_{\mu} = 0 \label{pi}
\end{equation}
Once we find the proper set of four null vectors with $\kappa = \epsilon = \pi = 0$, we can consider $l^{\mu}$ to be the propagation vector and $m^{\mu}$ to be the polarization vector. Along with $\kappa$, $\epsilon$ and $\pi$, two other \textit{spin coefficients} are relevant in the context of rotation and shear of an image. These spin coefficients are given by the following expressions.
\begin{subequations}
\begin{eqnarray}
\rho &=& m^{\mu}m^{*\nu}\nabla_{\nu}l_{\mu}, \label{rho} \\
\sigma &=& m^{\mu}m^{\nu}\nabla_{\nu}l_{\mu} \label{sigma}
\end{eqnarray}
\end{subequations}
The NP formalism also provides a systematic approach of achieving the condition, $\kappa = \epsilon = \pi = 0$. Initially, we have to find a set of null vectors with $\kappa = 0$ which can be done by solving the geodesic equations and exploiting Eqs.~(\ref{null})-(\ref{normalized}). Provided $\kappa = 0$, we can always make $\epsilon = 0$ by the rotation of class III. Under this rotation, the null vectors change in the following way.
\begin{eqnarray}
l^{\mu} \rightarrow A^{-1} l^{\mu},&& \;\;\; n^{\mu} \rightarrow A n^{\mu}, \nonumber \\
m^{\mu} \rightarrow e^{i\theta} m^{\mu},&& \;\;\; m^{*\mu} \rightarrow e^{-i\theta}m^{*\mu} \label{vec3}
\end{eqnarray}
where $A$ and $\theta$ are two real functions. Under this rotation, the relevant spin-coefficients change as the following.
\begin{eqnarray}
\kappa &&\rightarrow A^{-2} e^{i\theta}\kappa,\;\; \pi \rightarrow e^{-i\theta}\pi, \nonumber \\
\epsilon &&\rightarrow \frac{1}{A} \epsilon - \frac{1}{2A^{2}} l^{\mu}\partial_{\mu}A + \frac{i}{2A}l^{\mu}\partial_{\mu}\theta, \nonumber \\
\rho &&\rightarrow \frac{1}{A}\rho, \;\; \sigma \rightarrow \frac{1}{A}e^{2i\theta}\sigma \label{spin3}
\end{eqnarray}
So the rotation of class III does not affect the direction of $l^{\mu}$ and the vanishing of $\kappa$, but helps to find $A$ and $\theta$ such that the new $\epsilon$ vanishes. Once we have $\kappa = \epsilon = 0$, we can make $\pi = 0$ by rotation of class I. Under this rotation, the null vectors change as the following.
\begin{eqnarray}
l^{\mu} &&\rightarrow l^{\mu}, \; m^{\mu} \rightarrow m^{\mu} + zl^{\mu}, \; m^{*\mu} \rightarrow m^{*\mu} + z^{*}l^{\mu}, \nonumber \\
n^{\mu} &&\rightarrow n^{\mu} + z^{*}m^{\mu} + zm^{*\mu} + zz^{*}l^{\mu} \label{vec1}
\end{eqnarray}
where $z$ is a complex function. The relevant spin-coefficients change as the following.
\begin{eqnarray}
\kappa &&\rightarrow \kappa, \; \epsilon \rightarrow \epsilon + z^{*}\kappa, \nonumber \\
\pi &&\rightarrow \pi + 2z^{*}\epsilon + (z^{*})^{2}\kappa + l^{\mu}\partial_{\mu}z^{*}, \nonumber \\
\rho &&\rightarrow \rho + z^{*}\kappa, \; \sigma \rightarrow \sigma + z\kappa \label{spin1}
\end{eqnarray}
The rotation of class I does not affect $l^{\mu}$ and the vanishing of $\kappa$ and $\epsilon$, but it helps to find $z$ such that the new $\pi$ vanishes. One should also note that, if $\kappa = 0$, this rotation does not alter the values of $\rho$ and $\sigma$. At this stage, the construction of the proper set of null vectors is complete.

\section{Euler-Lagrange Equations for Perturbed FRW Metric}

The Lagrangian of a free particle in the perturbed FRW metric is given by the following.
\begin{equation}
L = \frac{1}{2} (1 + 2 \Psi) \dot{t}^{2} - \frac{1}{2}  a^{2}(t) (1 - 2 \Phi) \big (\dot{r}^{2} + r^{2} \dot{\theta}^{2} + r^{2} \sin^{2} \theta \dot{\phi}^{2} \big ) \label{lagperturbed}
\end{equation}
where the dot implies a derivative with respect to the affine parameter($\lambda$). In the zeroth order($\Psi = \Phi = 0$), we had the following solutions.
\begin{equation}
\dot{t} = \frac{1}{a}, \qquad \dot{r} = \frac{1}{a^{2}}, \qquad \dot{\theta} = 0 \label{trtheta}
\end{equation}
With these solutions; Im $\rho = 0$, $\sigma = 0$. Hence, the unperturbed FRW metric is free from both shear and rotation, as expected. Now we assume perturbations to these zeroth order solutions.
\begin{subequations}
\begin{eqnarray}
\dot{t} &&= \frac{1}{a} + \epsilon_{t} (t,r,\theta), \label{epsilont}\\
\dot{r} &&= \frac{1}{a^{2}} + \epsilon_{r} (t,r,\theta), \label{epsilonr} \\
\dot{\theta} &&= \epsilon_{\theta} (t,r,\theta) \label{epsilontheta}
\end{eqnarray}
\end{subequations}
where $\epsilon_{t}, \epsilon_{r}, \epsilon_{\theta}$ are small quantities and we will keep the terms that are linear in $\epsilon$'s. Now we will write down Euler-Lagrange equations for $t, r, \theta$ and $\phi$. For radial null geodesics, we set $\dot{\phi} = 0$. Then the three equations corresponding to $t, r, \theta$ take the following forms.
\begin{subequations}
\begin{eqnarray}
\dot{\epsilon_{t}} - \frac{H}{a} \epsilon_{t} - \frac{2H}{a^{2}} (\Psi + \Phi) + \frac{1}{a^{2}} \partial_{t} (\Psi - \Phi) + \frac{2}{a^{3}} \partial_{r}\Psi + 2 H \epsilon_{r} &&= 0, \label{epsilonteqn} \\
a^{2} \dot{\epsilon_{r}} + 2 a H \epsilon_{r} + \frac{1}{a^{2}} \partial_{r} (\Psi - \Phi) - \frac{2}{a} \partial_{t} \Phi &&= 0, \label{epsilonreqn} \\
a^{2} r^{2} \dot{\epsilon_{\theta}} + 2 r (1 + a H r) \epsilon_{\theta} + \frac{1}{a^{2}} \partial_{\theta} (\Psi + \Phi) &&= 0 \label{epsilonthetaeqn}
\end{eqnarray}
\end{subequations}
At this point, we should note that $\ddot{\theta} = \dot{\epsilon_{\theta}}$ is non-zero even if $\dot{\theta} = \epsilon_{\theta} = 0$. Therefore, $\dot{\theta}$ changes to some non-zero value even if it is set to be zero initially. For null geodesics, the Lagrangian is zero which implies the following.
\begin{equation}
\epsilon_{r} = \frac{\epsilon_{t}}{a} + \frac{( \Psi + \Phi )}{a^{2}} \label{epsilonrt}
\end{equation}
If we replace this expression of $\epsilon_{r}$ in Eq.~(\ref{epsilonreqn}), we get Eq.~(\ref{epsilonteqn}) which is expected. Writing $\epsilon_{r}$ in terms of $\epsilon_{t}$ in Eq.~(\ref{epsilonteqn}), we get
\begin{equation}
\dot{\epsilon_{t}} + \frac{H}{a} \epsilon_{t} + \frac{1}{a^{2}} \partial_{t} (\Psi - \Phi) + \frac{2}{a^{3}} \partial_{r} \Psi = 0 \label{epsilontonly}
\end{equation}

\begin{acknowledgments}
I am grateful to Pankaj Jain for suggesting this problem to me and for many enlightening discussions. I would also like to thank Pierre Sikivie, Steven Detweiler, Prabhakar Tiwari for useful discussions and to the anonymous referee for useful comments.
\end{acknowledgments}

\bibliography{Chakrabarty_bibliography}

\end{document}